\documentclass[11pt]{article}

\usepackage[preprint]{acl}

\usepackage{times}
\usepackage{latexsym}
\usepackage[T1]{fontenc}
\usepackage[utf8]{inputenc}
\usepackage{microtype}
\usepackage{inconsolata}
\usepackage{graphicx}

\usepackage{booktabs}
\usepackage{amsmath}
\usepackage{amsfonts}
\usepackage{xcolor}
\usepackage{colortbl}
\usepackage{multirow}
\usepackage{subcaption}
\usepackage{enumitem}
\usepackage{pgfplots}
\pgfplotsset{compat=1.17}
\usepackage{tikz}
\usetikzlibrary{trees,arrows.meta,positioning,shapes.geometric,calc,patterns}

\definecolor{pertcolor}{HTML}{1f77b4}
\definecolor{poiscolor}{HTML}{ff7f0e}
\definecolor{thrtcolor}{HTML}{2ca02c}
\definecolor{basecolor}{HTML}{7f7f7f}
\definecolor{mathred}{HTML}{d62728}
\definecolor{codeblu}{HTML}{1f77b4}
\definecolor{mtbpurp}{HTML}{9467bd}
\definecolor{lightgray}{HTML}{f0f0f0}

\title{DistillGuard: Evaluating Defenses Against LLM Knowledge Distillation}

\author{Bo Jiang\thanks{Corresponding author. Code and data will be released at \url{https://github.com/bojiang/distillguard} upon publication.} \\
  \texttt{bo.jiang@temple.edu}}

\begin{document}
\maketitle
\begin{abstract}
Knowledge distillation from proprietary LLM APIs poses a growing threat to model providers, yet defenses against this attack remain fragmented and unevaluated. We present DistillGuard, a framework for systematically evaluating output-level defenses against LLM knowledge distillation. We introduce a taxonomy of three defense categories---output perturbation, data poisoning, and information throttling---and evaluate nine defense configurations using a standardized pipeline with Qwen3-14B as teacher and Qwen2.5-7B-Instruct as student across three benchmarks (MATH-500, HumanEval+, MT-Bench). Our results reveal that, in a same-family distillation setting against a naive attacker, most output-level defenses are surprisingly ineffective: paraphrasing-based perturbation barely degrades distilled student quality, and data poisoning primarily impairs conversational fluency while leaving task-specific capabilities intact. Only chain-of-thought removal substantially impairs mathematical reasoning (31.4\% vs.\ 67.8\% baseline), though code generation remains unaffected. These findings demonstrate that the effectiveness of distillation defenses is highly task-dependent and that current output-level approaches are insufficient to broadly prevent knowledge theft.
\end{abstract}

\section{Introduction}
\label{sec:intro}

The rise of proprietary large language models (LLMs) accessed through APIs has created a lucrative ecosystem for model providers \citep{openai2024tos}. However, it has simultaneously exposed these models to \emph{knowledge distillation attacks}: an adversary queries the API with carefully chosen prompts, collects the responses, and uses them to train a smaller, cheaper student model that approximates the proprietary model's capabilities \citep{hinton2015distilling,xu2024survey}. This threat is not hypothetical---projects such as Stanford Alpaca \citep{taori2023alpaca}, Vicuna \citep{chiang2023vicuna}, and WizardLM \citep{xu2023wizardlm} have demonstrated that small open-source models can acquire substantial capabilities by training on outputs from proprietary APIs.

The economic implications are significant. When an attacker can replicate a model's capabilities through a few thousand API queries costing tens of dollars, the provider's investment in data curation, RLHF, and infrastructure is effectively expropriated. This has led providers to explicitly prohibit distillation in their terms of service \citep{openai2024tos}, but technical enforcement remains challenging: from the API's perspective, a distillation query is indistinguishable from a legitimate one.

Despite the severity of this threat, the landscape of defenses remains ad hoc. API providers have explored various countermeasures---from output paraphrasing and response truncation to deliberate data poisoning---but these defenses have been proposed and deployed in isolation, without systematic evaluation of their effectiveness. A provider considering whether to deploy output perturbation has no principled way to assess how much it will actually impair an attacker's distillation quality, or whether the defense's collateral damage to legitimate users justifies its protective benefit.

We address this gap with \textbf{DistillGuard}, a framework for systematically evaluating output-level defenses against LLM knowledge distillation. Our contributions are:

\begin{enumerate}[leftmargin=*]
  \item \textbf{Defense taxonomy.} We organize output-level defenses into three categories---perturbation, poisoning, and throttling---with representative implementations in each (Section~\ref{sec:taxonomy}).

  \item \textbf{Evaluation framework.} We define a standardized pipeline with a formal threat model and metrics for both distillation effectiveness (DE) and distillation cost (DC) (Section~\ref{sec:framework}).

  \item \textbf{Empirical evaluation and analysis.} We evaluate nine defense configurations across three benchmarks, revealing that most output-level defenses provide limited protection against even the simplest attacker. We identify cross-cutting patterns---including the empirical ineffectiveness of semantic-preserving perturbation and a task-dependency structure for information throttling---and analyze the cost-effectiveness trade-off of each defense (Sections~\ref{sec:results}--\ref{sec:analysis}).
\end{enumerate}

Our key finding is that output-level defenses are generally insufficient to prevent knowledge distillation. The most effective defense---chain-of-thought removal---achieves its impact by withholding reasoning traces rather than corrupting the output, and its effect is limited to reasoning-dependent tasks. These results suggest that providers seeking robust protection must look beyond output-level interventions toward structural defenses such as watermarking \citep{kirchenbauer2023watermark} or architectural safeguards.

\section{A Taxonomy of Distillation Defenses}
\label{sec:taxonomy}

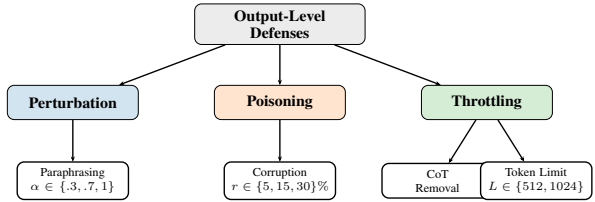
\begin{figure}[t]
\centering
\resizebox{\columnwidth}{!}{%
\begin{tikzpicture}[
  level 1/.style={sibling distance=42mm, level distance=16mm},
  level 2/.style={sibling distance=20mm, level distance=16mm},
  every node/.style={font=\small},
  root/.style={draw, rounded corners, fill=gray!15, font=\small\bfseries, text width=32mm, align=center, minimum height=7mm},
  cat/.style={draw, rounded corners, font=\small\bfseries, text width=24mm, align=center, minimum height=7mm},
  leaf/.style={draw, rounded corners, fill=white, font=\scriptsize, text width=20mm, align=center, minimum height=6mm},
  edge from parent/.style={draw, -{Stealth[length=3pt]}, thick},
]
\node[root] {Output-Level\\Defenses}
  child {
    node[cat, fill=pertcolor!20] {Perturbation}
    child { node[leaf] {Paraphrasing\\$\alpha \in \{.3,.7,1\}$} }
  }
  child {
    node[cat, fill=poiscolor!20] {Poisoning}
    child { node[leaf] {Corruption\\$r \in \{5,15,30\}\%$} }
  }
  child {
    node[cat, fill=thrtcolor!20] {Throttling}
    child { node[leaf] {CoT\\Removal} }
    child { node[leaf] {Token Limit\\$L \in \{512,1024\}$} }
  };
\end{tikzpicture}%
}
\caption{Taxonomy of output-level distillation defenses evaluated in this work. Each category targets a different mechanism: perturbation corrupts the signal, poisoning introduces adversarial errors, and throttling restricts information content.}
\label{fig:taxonomy}
\end{figure}

We organize output-level defenses against LLM distillation into three categories based on their mechanism of action (Figure~\ref{fig:taxonomy}). This taxonomy focuses on defenses that operate on the API's \emph{output stream}---they modify, corrupt, or restrict the teacher's response before it reaches the caller. We exclude input-level defenses (e.g., query detection, rate limiting) and model-level defenses (e.g., watermarking, differential privacy), which are orthogonal and can be combined with output-level approaches.

\subsection{Output Perturbation}

Output perturbation defenses modify the teacher's response while attempting to preserve its approximate meaning. The goal is to inject noise that degrades the distillation signal without making the response useless to legitimate users.

We implement \emph{paraphrasing-based perturbation}. Given a teacher response $y$, a separate paraphraser model $P$ rewrites it with controllable perturbation strength $\alpha \in [0, 1]$:
\begin{equation}
  \tilde{y} = P(y, \alpha)
\end{equation}

The strength parameter $\alpha$ is communicated to the paraphraser through its system prompt: at $\alpha = 0$ the paraphraser is instructed to preserve the response exactly, while at $\alpha = 1.0$ it is given maximum freedom to restructure, rephrase, and reorganize the content while maintaining correctness.  We evaluate $\alpha \in \{0.3, 0.7, 1.0\}$ to span the range from conservative to aggressive paraphrasing.

The intuition behind this defense is that paraphrasing introduces stylistic and structural variation that prevents the student from learning the teacher's precise output distribution. If the attacker collects paraphrased responses, they are effectively training on a noisier version of the teacher's distribution.

\subsection{Data Poisoning}

Data poisoning defenses deliberately inject incorrect information into a fraction of responses. Rather than adding noise to the signal, they inject adversarial errors that the student may internalize.

We implement \emph{response corruption}: for a randomly selected fraction $r$ of prompts, the teacher's correct response is replaced by a plausible but intentionally incorrect one. The incorrect responses are generated by prompting the teacher model itself to ``solve the problem but arrive at the wrong answer,'' ensuring that the poisoned responses are stylistically consistent with genuine ones and not trivially detectable. We evaluate $r \in \{0.05, 0.15, 0.30\}$.

The strength of this defense is its subtlety: unlike perturbation, which merely adds noise, poisoning introduces systematically wrong training signal that may propagate through the student's learned representations. Its weakness is that it degrades the API's output quality for legitimate users in proportion to the poison rate.

\subsection{Information Throttling}

Information throttling defenses restrict the \emph{information content} of the teacher's response rather than corrupting it. The key insight is that the richness of the response---particularly the presence of chain-of-thought reasoning traces \citep{wei2022chain}---provides a strong supervisory signal for distillation \citep{hsieh2023distilling,mukherjee2023orca}. By limiting this information, the defense aims to reduce what the student can learn.

We implement two variants:

\paragraph{Chain-of-thought (CoT) removal.} The defense strips the reasoning trace from the teacher's response, returning only the final answer. For a math problem, this means returning ``42'' rather than the multi-step derivation. For code, this means returning only the function body without explanatory comments.

\paragraph{Token truncation.} The defense limits responses to a maximum of $L$ tokens. We evaluate $L \in \{512, 1024\}$, compared to the teacher's unconstrained generation of up to 4{,}096 tokens. Truncation is a blunter instrument than CoT removal---it may cut off important content regardless of whether it constitutes reasoning traces or final answers.

\subsection{Excluded Defense Categories}

Our taxonomy deliberately excludes several defense categories that we plan to evaluate in future work:

\begin{itemize}[leftmargin=*]
  \item \textbf{Adversarial decoding strategies (ADS)}: Modifying the teacher's sampling procedure (e.g., temperature, nucleus parameters) to produce outputs that are correct but harder to distill from.
  \item \textbf{Watermarking} \citep{kirchenbauer2023watermark}: Embedding statistical patterns in the token distribution that enable post-hoc detection but do not degrade quality.
  \item \textbf{Query detection}: Input-level defenses that identify and reject suspected distillation queries based on query patterns or rate limits.
\end{itemize}

These categories operate through different mechanisms (generation-time, model-level, or input-level) and require different evaluation methodologies. We focus on output-level post-processing defenses as the most commonly discussed and easiest to deploy.

\section{Evaluation Framework}
\label{sec:framework}

\subsection{Threat Model}
\label{sec:threat}

We consider a threat model with the following participants and assumptions:

\paragraph{Model provider.} The provider serves a proprietary teacher model $T$ through an API. The provider can apply output-level defenses---modifying, corrupting, or truncating responses before returning them---but cannot inspect the attacker's downstream use of the responses. The provider's goal is to minimize the quality of any student model trained on API outputs while maintaining service quality for legitimate users.

\paragraph{Attacker.} The attacker has access to (1)~the provider's API, which they can query with arbitrary prompts; (2)~a base student model $S$ (not necessarily related to the teacher); and (3)~compute resources for supervised fine-tuning. The attacker's goal is to maximize the student's quality on target benchmarks. In this paper, we evaluate the simplest attacker: a \emph{naive attacker} who queries each prompt once, collects the response verbatim, and trains the student on the resulting dataset without any filtering or post-processing.

\paragraph{Scope.} We evaluate defenses against the naive attacker as a lower bound on defense effectiveness. More sophisticated attackers---such as consensus voting (querying $k$ times and synthesizing), answer verification (filtering incorrect responses), or defense-aware adaptation---can only improve upon the naive attacker's results, meaning any defense that fails against the naive attacker will also fail against stronger attackers.

\subsection{The ProtectedAPI Abstraction}

We model the provider's defended system as a \emph{ProtectedAPI} that wraps the teacher model with a configurable defense pipeline. Given an input prompt $x$, the ProtectedAPI produces a defended response through three stages:

\begin{enumerate}[leftmargin=*]
  \item \textbf{Generation.} The teacher model generates a response: $y = T(x)$.
  \item \textbf{Defense.} A defense function $D$ transforms the response: $\tilde{y} = D(y, x)$. For perturbation, $D$ paraphrases; for poisoning, $D$ occasionally replaces with an incorrect response; for throttling, $D$ truncates or strips reasoning.
  \item \textbf{Return.} The defended response $\tilde{y}$ is returned to the caller.
\end{enumerate}

This abstraction ensures that all experiments use identical teacher outputs, with only the defense function varying. It also enables composing multiple defenses in sequence, though we evaluate only single defenses in this work.

\subsection{Distillation Pipeline}

Each experiment follows a fixed four-stage pipeline:
\begin{enumerate}[leftmargin=*]
  \item \textbf{Teacher generation.} The teacher generates responses for all 10{,}000 training prompts. These undefended responses are cached and reused across all experiments.
  \item \textbf{Defense application.} The defense function $D$ is applied to each teacher response, producing the defended training set $\{(x_i, \tilde{y}_i)\}_{i=1}^{N}$.
  \item \textbf{Student training.} The student model is fine-tuned on the defended training set via LoRA SFT \citep{hu2022lora}.
  \item \textbf{Evaluation.} The trained student is evaluated on three held-out benchmarks.
\end{enumerate}

By caching teacher responses in stage~1, we ensure that observed differences between experiments are attributable solely to the defense, not to stochastic variation in teacher generation.

\subsection{Metrics}

We define two complementary metrics that capture both sides of the defense trade-off.

\paragraph{Distillation effectiveness (DE).}
DE measures how well the distilled student retains quality under a defense. For each benchmark $b$:
\begin{equation}
  \text{DE}_b = \frac{S_b(\text{student}_\text{defended})}{S_b(\text{student}_\text{baseline})}
\label{eq:de}
\end{equation}
where $S_b(\cdot)$ denotes the score on benchmark $b$. A value of $\text{DE}_b = 1$ indicates the defense has no protective effect, while $\text{DE}_b \ll 1$ indicates strong protection. From the defender's perspective, an effective defense minimizes DE across all benchmarks.

We also define an aggregate DE across benchmarks:
\begin{equation}
  \overline{\text{DE}} = \frac{1}{|\mathcal{B}|} \sum_{b \in \mathcal{B}} \text{DE}_b
\label{eq:de_avg}
\end{equation}

\paragraph{Distillation cost (DC).}
DC measures the collateral damage to legitimate users---how much the defense degrades the quality of the API's own output:
\begin{equation}
  \text{DC}_b = 1 - \frac{S_b(\text{teacher}_\text{defended})}{S_b(\text{teacher}_\text{undefended})}
\label{eq:dc}
\end{equation}

A value of $\text{DC}_b = 0$ means no degradation to the user experience; $\text{DC}_b \to 1$ means the defense renders the API useless. An ideal defense achieves low DE (strong protection) with low DC (minimal user impact).

We evaluate DC by running the defended \emph{teacher} outputs through the same three benchmarks used for student evaluation (MATH-500, HumanEval+, MT-Bench). The teacher scores without any defense serve as the reference, and DC captures the proportional degradation introduced by each defense. We report DC alongside DE in our main results (Table~\ref{tab:results}) and analyze the cost-effectiveness trade-off in Section~\ref{sec:dc_analysis}.

\section{Experimental Setup}
\label{sec:setup}

\subsection{Models}

\paragraph{Teacher.} We use \textbf{Qwen3-14B} \citep{qwen2024qwen2} in non-thinking mode (chain-of-thought generation disabled at the model level) as the teacher. This represents a mid-size proprietary model served through an API. Responses are generated with greedy decoding (temperature 0) and a maximum length of 4{,}096 tokens.

\paragraph{Student.} We use \textbf{Qwen2.5-7B-Instruct} \citep{qwen2024qwen2} as the student base model. This represents an attacker who starts with a capable open-source instruction-tuned model and seeks to further enhance it through distillation. We deliberately choose an already-capable base model to reflect the realistic scenario where the attacker's starting point is not a blank slate.

\paragraph{Paraphraser.} For the perturbation defense, we use a separate Qwen2.5-7B-Instruct instance as the paraphraser. Using a separate model (rather than the teacher itself) reflects the practical constraint that the defense pipeline should not require additional forward passes of the expensive teacher model.

\subsection{Training Data}

We construct a set of 10{,}000 training prompts spanning three domains:
\begin{itemize}[leftmargin=*]
  \item \textbf{Mathematical reasoning} (3{,}000 prompts): Sampled from the MATH dataset \citep{hendrycks2021math} at difficulty levels 3--5, covering algebra, number theory, geometry, combinatorics, and calculus.
  \item \textbf{Code generation} (3{,}000 prompts): Programming challenges requiring function implementation in Python, covering algorithms, data structures, and string manipulation.
  \item \textbf{Open-ended instruction following} (4{,}000 prompts): Diverse prompts including creative writing, explanation, summarization, and multi-step reasoning tasks.
\end{itemize}

This domain distribution ensures that our evaluation captures defense effects across multiple capability dimensions. The math and code prompts have verifiable correct answers, enabling precise measurement, while open-ended prompts test the harder-to-quantify aspects of language model quality.

\subsection{Training Configuration}

All student models are fine-tuned with LoRA \citep{hu2022lora} using identical hyperparameters:
\begin{itemize}[leftmargin=*]
  \item LoRA rank $r = 16$, scaling factor $\alpha = 32$, dropout 0.05
  \item Target modules: all attention projections ($\mathbf{W}_q$, $\mathbf{W}_k$, $\mathbf{W}_v$, $\mathbf{W}_o$) and MLP projections ($\mathbf{W}_\text{gate}$, $\mathbf{W}_\text{up}$, $\mathbf{W}_\text{down}$)
  \item Optimizer: AdamW \citep{loshchilov2019adamw}, lr = $2 \times 10^{-4}$, weight decay 0.01
  \item Batch size 4, gradient accumulation 4 (effective batch 16)
  \item 3 epochs, 3\% linear warmup, maximum sequence length 4{,}096
  \item bfloat16 mixed precision
\end{itemize}

Using identical training configurations across all experiments is critical: it ensures that observed quality differences are attributable to the defense rather than to training hyperparameter choices.

\subsection{Evaluation Benchmarks}

We evaluate on three complementary benchmarks that collectively cover mathematical reasoning, code generation, and general instruction following:

\paragraph{MATH-500} \citep{hendrycks2021math}: 500 competition mathematics problems spanning seven topics. Evaluation uses exact match with SymPy-based symbolic equivalence checking to handle algebraic equivalences (e.g., $\frac{1}{2}$ vs.\ $0.5$). This benchmark tests the student's ability to perform multi-step mathematical reasoning---a capability known to benefit strongly from chain-of-thought training data \citep{wei2022chain}.

\paragraph{HumanEval+} \citep{liu2024evalplus}: 164 Python programming problems from the original HumanEval benchmark \citep{chen2021evaluating} augmented with additional test cases to reduce false positives. Each problem provides a function signature and docstring; the model must generate a correct implementation. This benchmark tests code generation---a capability where the output itself embodies the reasoning process.

\paragraph{MT-Bench} \citep{zheng2023judging}: 80 multi-turn questions spanning 8 categories (writing, roleplay, reasoning, math, coding, extraction, STEM, humanities), scored 1--10 by an LLM judge. Unlike MATH-500 and HumanEval+, which have objectively verifiable answers, MT-Bench evaluates open-ended response quality, fluency, and instruction following. We use Qwen2.5-7B-Instruct as the judge model for consistency.

\subsection{Experiment Grid}

Table~\ref{tab:experiments} summarizes the ten experiments. Experiment A01 is the no-defense baseline; A02--A10 evaluate nine defense configurations spanning three categories with varying strength parameters.

\begin{table}[t]
\centering
\small
\begin{tabular}{llll}
\toprule
\textbf{ID} & \textbf{Category} & \textbf{Defense} & \textbf{Param.} \\
\midrule
A01 & --- & No defense & --- \\
\midrule
A02 & Perturbation & Paraphrase & $\alpha = 0.3$ \\
A03 & Perturbation & Paraphrase & $\alpha = 0.7$ \\
A04 & Perturbation & Paraphrase & $\alpha = 1.0$ \\
\midrule
A05 & Poisoning & Corruption & $r = 5\%$ \\
A06 & Poisoning & Corruption & $r = 15\%$ \\
A07 & Poisoning & Corruption & $r = 30\%$ \\
\midrule
A08 & Throttling & CoT removal & strip reasoning \\
A09 & Throttling & Token limit & $L = 512$ \\
A10 & Throttling & Token limit & $L = 1024$ \\
\bottomrule
\end{tabular}
\caption{Experiment grid: nine defense configurations spanning three categories with increasing strength.}
\label{tab:experiments}
\end{table}

\section{Results}
\label{sec:results}

\subsection{Main Results}

Table~\ref{tab:results} presents the full results across all ten experiments, reporting both raw scores and distillation effectiveness (DE) relative to the baseline.

\begin{table*}[t]
\centering
\begin{tabular}{ll ccc ccc cc}
\toprule
& & \multicolumn{3}{c}{\textbf{Raw Scores}} & \multicolumn{3}{c}{\textbf{Distillation Effectiveness (DE)}} & & \\
\cmidrule(lr){3-5} \cmidrule(lr){6-8}
\textbf{ID} & \textbf{Defense} & \textbf{MATH} & \textbf{HE+} & \textbf{MT-B} & \textbf{MATH} & \textbf{HE+} & \textbf{MT-B} & $\overline{\textbf{DE}}$ & $\overline{\textbf{DC}}$ \\
\midrule
A01 & \textit{Baseline (no defense)} & 67.8 & 91.5 & 8.20 & 1.000 & 1.000 & 1.000 & 1.000 & --- \\
\midrule
\rowcolor{pertcolor!8}
A02 & Paraphrase $\alpha{=}0.3$ & 66.0 & 92.7 & 8.08 & 0.973 & 1.013 & 0.985 & 0.990 & 0.030 \\
\rowcolor{pertcolor!8}
A03 & Paraphrase $\alpha{=}0.7$ & 63.4 & 90.9 & 8.24 & 0.935 & 0.993 & 1.005 & 0.978 & 0.034 \\
\rowcolor{pertcolor!8}
A04 & Paraphrase $\alpha{=}1.0$ & 69.4 & 90.2 & 8.41 & 1.024 & 0.986 & 1.026 & 1.012 & 0.070 \\
\midrule
\rowcolor{poiscolor!8}
A05 & Poison $r{=}5\%$ & 66.6 & 91.5 & 7.84 & 0.982 & 1.000 & 0.956 & 0.979 & 0.017 \\
\rowcolor{poiscolor!8}
A06 & Poison $r{=}15\%$ & 65.6 & 89.0 & 7.75 & 0.968 & 0.973 & 0.945 & 0.962 & 0.052 \\
\rowcolor{poiscolor!8}
A07 & Poison $r{=}30\%$ & 66.6 & 93.9 & 7.63 & 0.982 & 1.026 & 0.930 & 0.980 & 0.098 \\
\midrule
\rowcolor{thrtcolor!8}
A08 & CoT removal & 31.4 & 93.9 & 7.74 & \textbf{0.463} & 1.026 & 0.944 & 0.811 & \textbf{0.311} \\
\rowcolor{thrtcolor!8}
A09 & Token limit $L{=}512$ & 64.4 & 93.3 & 8.10 & 0.950 & 1.020 & 0.988 & 0.986 & 0.046 \\
\rowcolor{thrtcolor!8}
A10 & Token limit $L{=}1024$ & 67.4 & 89.0 & 8.11 & 0.994 & 0.973 & 0.989 & 0.985 & 0.014 \\
\bottomrule
\end{tabular}
\caption{Full results for all ten experiments. Raw scores: MATH-500 accuracy (\%), HumanEval+ pass rate (\%), MT-Bench score (1--10). DE is the ratio to baseline (Eq.~\ref{eq:de}); $\overline{\text{DE}}$ is the average across benchmarks. $\overline{\text{DC}}$ is the average distillation cost (Eq.~\ref{eq:dc})---how much the defense degrades the teacher's own output for legitimate users. An ideal defense has low DE and low DC. Bold marks notable outliers.}
\label{tab:results}
\end{table*}

The most striking finding is that \textbf{the vast majority of DE values are close to~1.0}. Of the 27 benchmark--defense combinations (9 defenses $\times$ 3 benchmarks), only one---CoT removal on MATH-500 ($\text{DE} = 0.463$)---shows a substantial protective effect. All other DE values fall in the range $[0.930, 1.026]$, indicating that the defenses barely impair distillation quality. In several cases (A04, A07, A08 on HumanEval+), the defended student \emph{outperforms} the baseline, suggesting that certain defense-induced perturbations can act as beneficial regularization. We note that these results are obtained in a same-family distillation setting (Qwen3-14B $\to$ Qwen2.5-7B-Instruct) with a naive attacker; cross-family settings or adaptive attackers may yield different dynamics (see Section~\ref{sec:limitations}).

\subsection{Output Perturbation Is Ineffective}

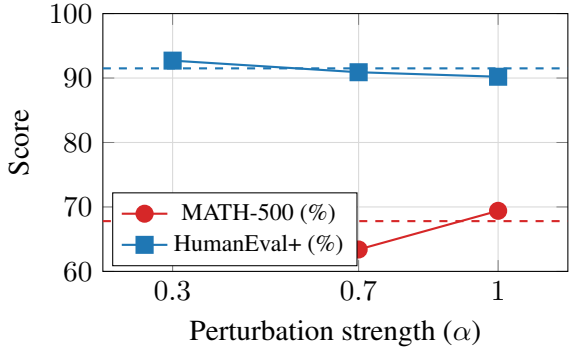
\begin{figure}[t]
\centering
\begin{tikzpicture}
\begin{axis}[
  width=\columnwidth, height=5cm,
  xlabel={Perturbation strength ($\alpha$)},
  ylabel={Score},
  xmin=0.15, xmax=1.15,
  ymin=60, ymax=100,
  xtick={0.3, 0.7, 1.0},
  legend style={at={(0.02,0.02)}, anchor=south west, font=\small},
  grid=major, grid style={gray!30},
  every axis plot/.append style={thick, mark size=3pt},
]
\addplot[color=mathred, mark=*] coordinates {(0.3,66.0) (0.7,63.4) (1.0,69.4)};
\addplot[color=codeblu, mark=square*] coordinates {(0.3,92.7) (0.7,90.9) (1.0,90.2)};
\addplot[color=mathred, dashed, forget plot] coordinates {(0.15,67.8) (1.15,67.8)};
\addplot[color=codeblu, dashed, forget plot] coordinates {(0.15,91.5) (1.15,91.5)};
\legend{MATH-500 (\%), HumanEval+ (\%)}
\end{axis}
\end{tikzpicture}
\caption{Perturbation strength vs.\ student quality. Dashed lines show the baseline. No consistent degradation occurs at any perturbation level; $\alpha = 1.0$ slightly \emph{improves} MATH performance.}
\label{fig:perturbation}
\end{figure}

Paraphrasing-based perturbation (A02--A04) provides essentially no protection across any benchmark or strength level (Figure~\ref{fig:perturbation}). Even at maximum perturbation strength ($\alpha = 1.0$), the distilled student achieves 69.4\% on MATH-500---\emph{higher} than the undefended baseline of 67.8\%. The per-benchmark DE values across all three strength levels remain within $\pm 3\%$ of 1.0, and the aggregate $\overline{\text{DE}}$ ranges from 0.978 to 1.012.

Notably, there is no monotonic relationship between perturbation strength and quality degradation: the strongest perturbation ($\alpha = 1.0$) yields the best math score among perturbation variants, while moderate perturbation ($\alpha = 0.7$) yields the lowest. This non-monotonicity further undermines the premise of perturbation as a defense, suggesting that the observed variations are noise rather than a systematic defense effect.

\subsection{Poisoning Degrades Conversational Quality}

\begin{figure}[t]
\centering
\resizebox{\columnwidth}{!}{%
\begin{tikzpicture}
\begin{axis}[
  width=0.95\columnwidth, height=5cm,
  xlabel={Poison rate (\%)},
  ylabel={MATH-500 / HE+ (\%)},
  axis y line*=left,
  xmin=0, xmax=35,
  ymin=60, ymax=100,
  xtick={5, 15, 30},
  legend style={at={(0.02,0.02)}, anchor=south west, font=\small},
  grid=major, grid style={gray!30},
  every axis plot/.append style={thick, mark size=3pt},
]
\addplot[color=mathred, mark=*] coordinates {(5,66.6) (15,65.6) (30,66.6)};
\addplot[color=codeblu, mark=square*] coordinates {(5,91.5) (15,89.0) (30,93.9)};
\addplot[color=mathred, dashed, forget plot] coordinates {(0,67.8) (35,67.8)};
\legend{MATH-500, HumanEval+}
\end{axis}
\begin{axis}[
  width=0.95\columnwidth, height=5cm,
  axis y line*=right, axis x line=none,
  ylabel={MT-Bench},
  ylabel style={color=mtbpurp},
  xmin=0, xmax=35,
  ymin=7.4, ymax=8.6,
  ytick={7.5, 7.75, 8.0, 8.25},
  tick label style={color=mtbpurp},
  legend style={at={(0.98,0.02)}, anchor=south east, font=\small},
  every axis plot/.append style={thick, mark size=3pt},
]
\addplot[color=mtbpurp, mark=triangle*] coordinates {(5,7.84) (15,7.75) (30,7.63)};
\addplot[color=mtbpurp, dashed, forget plot] coordinates {(0,8.20) (35,8.20)};
\legend{MT-Bench}
\end{axis}
\end{tikzpicture}%
}
\caption{Poisoning rate vs.\ student quality. MT-Bench (right axis) degrades monotonically while MATH and HumanEval+ remain stable. Dashed lines show baselines.}
\label{fig:poisoning}
\end{figure}
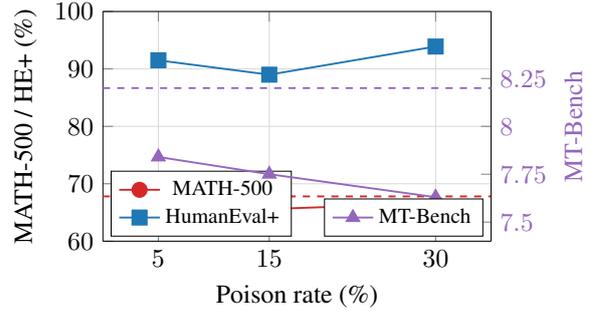

Data poisoning (A05--A07) exhibits a \emph{task-selective} degradation pattern (Figure~\ref{fig:poisoning}). The primary effect is on MT-Bench scores, which decrease monotonically from 8.20 (baseline) to 7.63 (30\% poison rate), representing a DE of 0.930. However, MATH-500 and HumanEval+ scores remain largely unaffected, with DE values between 0.968 and 1.026.

This asymmetry suggests that poisoning primarily corrupts the model's \emph{conversational behavior}---its ability to produce fluent, well-structured, and contextually appropriate responses---rather than its task-specific reasoning or coding capabilities. The poisoned training examples likely teach the student subtly incorrect response patterns that manifest most clearly in open-ended evaluation (MT-Bench) but do not interfere with the structured problem-solving required by math and code benchmarks.

An unexpected finding is that A07 (30\% poison) achieves the highest HumanEval+ score of 93.9\%, \emph{above} the undefended baseline. This may reflect a selection effect: when 30\% of training examples are corrupted, the student effectively learns from a smaller clean dataset, which may coincidentally reduce overfitting on code tasks. This finding also suggests that code capabilities are particularly robust to training data corruption.

\subsection{Information Throttling Is Task-Dependent}
\label{sec:throttling}

Information throttling produces the most varied results across our evaluation.

\paragraph{CoT removal (A08)} causes a dramatic collapse in mathematical reasoning: MATH-500 drops from 67.8\% to 31.4\%, a DE of 0.463---the only substantial protective effect observed in any experiment. However, this effect is entirely confined to math: HumanEval+ actually \emph{increases} to 93.9\% (DE = 1.026), and MT-Bench remains at 7.74 (DE = 0.944).

The magnitude of the math degradation is remarkable. The defended student's 31.4\% accuracy is substantially \emph{worse} than the base Qwen2.5-7B-Instruct model before any distillation, suggesting that training on answer-only data actively harms reasoning by teaching the student to skip intermediate steps. This is consistent with the finding that CoT traces are not merely helpful but \emph{necessary} for distilling mathematical reasoning \citep{hsieh2023distilling}.

\paragraph{Token truncation (A09, A10)} produces moderate and diminishing effects. At $L = 512$, MATH-500 drops to 64.4\% (DE = 0.950), a modest 3.4 percentage point decline. At $L = 1024$, the effect is negligible (DE = 0.994). Neither truncation level substantially affects HumanEval+ or MT-Bench, indicating that the critical information for code and conversation tasks is contained within the first 512 tokens.

\section{Analysis}
\label{sec:analysis}

\subsection{Summary by Defense Category}

Table~\ref{tab:summary} summarizes the average DE per defense category, revealing three distinct defense profiles.

\begin{table}[t]
\centering
\small
\begin{tabular}{lcccc}
\toprule
\textbf{Category} & \textbf{DE$_\text{MATH}$} & \textbf{DE$_\text{HE+}$} & \textbf{DE$_\text{MT-B}$} & $\overline{\textbf{DE}}$ \\
\midrule
Perturbation & 0.977 & 0.997 & 1.005 & 0.993 \\
Poisoning & 0.977 & 1.000 & 0.944 & 0.974 \\
Throttling & 0.802 & 1.006 & 0.974 & 0.927 \\
\bottomrule
\end{tabular}
\caption{Average DE by defense category. Only throttling substantially reduces $\overline{\text{DE}}$, driven almost entirely by CoT removal's effect on math.}
\label{tab:summary}
\end{table}

\begin{itemize}[leftmargin=*]
  \item \textbf{Perturbation} ($\overline{\text{DE}} = 0.993$): Uniformly ineffective across all dimensions.
  \item \textbf{Poisoning} ($\overline{\text{DE}} = 0.974$): Selectively impairs conversational quality while leaving task-specific skills intact.
  \item \textbf{Throttling} ($\overline{\text{DE}} = 0.927$): Dramatically impairs mathematical reasoning but not other capabilities; the average is heavily influenced by CoT removal.
\end{itemize}

No defense category achieves uniformly low DE across all benchmarks. This is the central negative result of our evaluation: \textbf{no existing output-level defense provides comprehensive protection against knowledge distillation}.

\subsection{On the Limitations of Semantic-Preserving Perturbation}

The complete failure of perturbation defense reveals what we term the \emph{perturbation limitation}: empirically, any semantic-preserving transformation of a correct response appears to preserve the distillation signal. We observe that if a defense $D$ satisfies two properties---
\begin{enumerate}[leftmargin=*]
  \item \emph{Correctness preservation}: $D(y)$ is a valid answer to the prompt whenever $y$ is.
  \item \emph{Semantic preservation}: $D(y)$ conveys approximately the same information as $y$.
\end{enumerate}
---then training on $\{(x_i, D(y_i))\}$ should yield a student comparable to training on $\{(x_i, y_i)\}$. Paraphrasing satisfies both properties by design, which explains its failure as a defense. We stress that this is an empirical observation within our experimental setting (same-family distillation, single attacker level), not a formal impossibility proof---different model families or distillation setups may exhibit different sensitivity to perturbation.

This limitation implies that effective output-level defenses must either sacrifice correctness (poisoning), sacrifice information content (throttling), or operate outside the output stream entirely (watermarking, query detection). The perturbation category is fundamentally constrained by the tension between preserving user value and degrading distillation value.

\subsection{Task-Defense Interaction Structure}

\begin{figure}[t]
\centering
\begin{tikzpicture}
\begin{axis}[
  ybar,
  width=0.95\columnwidth, height=5.5cm,
  bar width=7pt,
  xlabel={Defense Category},
  ylabel={Average DE},
  ymin=0.7, ymax=1.1,
  xtick=data,
  symbolic x coords={Perturbation, Poisoning, Throttling},
  legend style={at={(0.5,1.02)}, anchor=south, font=\small, legend columns=3},
  grid=major, grid style={gray!20},
  nodes near coords style={font=\tiny, rotate=90, anchor=west},
  every node near coord/.append style={xshift=0pt, yshift=2pt},
]
\addplot[fill=mathred!80, draw=mathred!90] coordinates
  {(Perturbation,0.977) (Poisoning,0.977) (Throttling,0.802)};
\addplot[fill=codeblu!80, draw=codeblu!90] coordinates
  {(Perturbation,0.997) (Poisoning,1.000) (Throttling,1.006)};
\addplot[fill=mtbpurp!80, draw=mtbpurp!90] coordinates
  {(Perturbation,1.005) (Poisoning,0.944) (Throttling,0.974)};
\legend{MATH-500, HumanEval+, MT-Bench}
\end{axis}
\end{tikzpicture}
\caption{Average DE by defense category and benchmark. Each category primarily affects a different capability: poisoning targets conversation, throttling targets reasoning, perturbation affects none.}
\label{fig:interaction}
\end{figure}
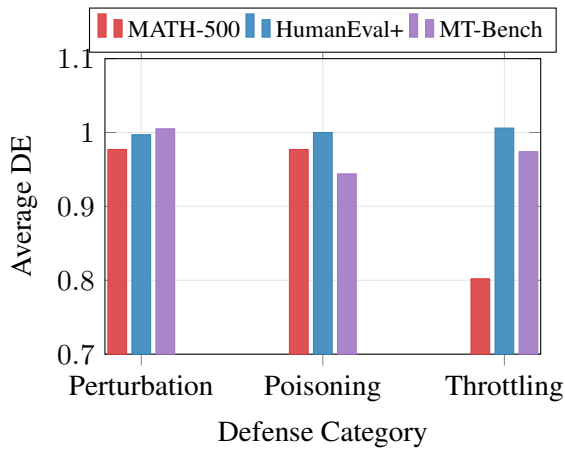

Our results reveal a structured interaction between defense type and task type, visualized in Figure~\ref{fig:interaction}.

Each defense category exhibits a characteristic ``damage profile'':
\begin{itemize}[leftmargin=*]
  \item Perturbation affects no capability dimension.
  \item Poisoning primarily damages \emph{open-ended response quality} (MT-Bench), presumably because poisoned examples corrupt response style and structure.
  \item Throttling primarily damages \emph{mathematical reasoning} (MATH-500), because reasoning traces are critical for learning decomposition skills.
  \item Code generation (HumanEval+) is remarkably \emph{robust to all defenses}, with DE consistently $\geq 0.973$.
\end{itemize}

The robustness of code generation is particularly noteworthy. We hypothesize that code has a built-in ``self-contained reasoning'' property: the code itself embodies the solution logic, so even without explicit reasoning traces, the student can learn correct solution patterns by imitating the code. Additionally, code is more constrained than natural language (it must parse and pass test cases), which may make it harder to corrupt through paraphrasing or poisoning.

\subsection{Defense Strength Scaling}

Within each defense category, we observe different scaling behaviors as defense strength increases:

\paragraph{Perturbation ($\alpha$).} No consistent scaling---the strongest perturbation ($\alpha = 1.0$) \emph{improves} student quality on math and MT-Bench. This confirms that perturbation strength is not a useful defense knob.

\paragraph{Poisoning ($r$).} Linear degradation on MT-Bench ($r = 0.05 \to 0.15 \to 0.30$ maps to MT-B $= 7.84 \to 7.75 \to 7.63$), but the slope is shallow: each 10\% increase in poison rate reduces the MT-Bench score by only $\sim$0.1 points. Extrapolating, even 100\% poisoning (which renders the API useless) would reduce MT-Bench by only $\sim$1 point. This suggests poisoning has diminishing returns even for its primary effect.

\paragraph{Truncation ($L$).} Threshold behavior on MATH-500: truncation to 1024 tokens has negligible effect (DE = 0.994), while truncation to 512 tokens produces a modest decline (DE = 0.950). This suggests that the marginal information content of tokens beyond $\sim$1024 is minimal for most responses, but the 512--1024 range contains some useful reasoning content.

\subsection{Defense Cost-Effectiveness}
\label{sec:dc_analysis}

A complete evaluation of defenses must account for their cost to legitimate users. Table~\ref{tab:dc} reports the per-benchmark distillation cost (DC) for each defense, computed from the teacher model (Qwen3-14B) evaluated with and without each defense applied.

\begin{table}[t]
\centering
\small
\begin{tabular}{llcccc}
\toprule
\textbf{ID} & \textbf{Defense} & \textbf{DC$_\text{MATH}$} & \textbf{DC$_\text{HE+}$} & \textbf{DC$_\text{MT-B}$} & $\overline{\textbf{DC}}$ \\
\midrule
\rowcolor{pertcolor!8}
A02 & Paraph.\ $\alpha{=}0.3$ & $\approx$0 & .093 & .000 & .030 \\
\rowcolor{pertcolor!8}
A03 & Paraph.\ $\alpha{=}0.7$ & .000 & .099 & .003 & .034 \\
\rowcolor{pertcolor!8}
A04 & Paraph.\ $\alpha{=}1.0$ & .079 & .117 & .015 & .070 \\
\midrule
\rowcolor{poiscolor!8}
A05 & Poison $r{=}5\%$ & .038 & .000 & .012 & .017 \\
\rowcolor{poiscolor!8}
A06 & Poison $r{=}15\%$ & .128 & .012 & .016 & .052 \\
\rowcolor{poiscolor!8}
A07 & Poison $r{=}30\%$ & .235 & .012 & .046 & .098 \\
\midrule
\rowcolor{thrtcolor!8}
A08 & CoT removal & \textbf{.839} & .074 & .021 & \textbf{.311} \\
\rowcolor{thrtcolor!8}
A09 & Limit $L{=}512$ & .130 & .000 & .009 & .046 \\
\rowcolor{thrtcolor!8}
A10 & Limit $L{=}1024$ & .041 & .000 & .002 & .014 \\
\bottomrule
\end{tabular}
\caption{Per-benchmark distillation cost (DC). DC = $1 - S_\text{defended}/S_\text{undefended}$ for the teacher model itself. Higher values mean greater degradation for legitimate users. Teacher reference scores: MATH-500 78.4\%, HumanEval+ 98.8\%, MT-Bench 8.44.}
\label{tab:dc}
\end{table}

Several patterns emerge from the DC analysis:

\paragraph{CoT removal paradox.} CoT removal (A08) is the only defense with substantial DE, but it also has the highest DC ($\overline{\text{DC}} = 0.311$), driven almost entirely by MATH-500 ($\text{DC}_\text{MATH} = 0.839$). Removing chain-of-thought traces devastates the teacher's \emph{own} math accuracy (from 78.4\% to 12.6\%), meaning the defense imposes a severe cost on legitimate users of mathematical reasoning. This creates a harsh trade-off: the defense works precisely \emph{because} it destroys the reasoning content, but this destruction hurts users as much as it hurts attackers.

\paragraph{Perturbation on code.} Paraphrasing incurs noticeable DC on HumanEval+ (DC$_\text{HE+}$ = 0.09--0.12) despite having negligible DC on math and MT-Bench. The paraphraser subtly corrupts code syntax during rewriting---introducing variable renames, restructured logic, or broken edge cases---even when instructed to preserve correctness. Combined with perturbation's near-zero DE, this makes it the worst defense profile: it hurts legitimate users without protecting against distillation.

\paragraph{Poisoning scales linearly.} Poisoning DC on MATH-500 scales approximately linearly with the poison rate: DC$_\text{MATH} \approx 0.8r$ (0.038, 0.128, 0.235 for $r$ = 0.05, 0.15, 0.30). This is expected: a fraction $r$ of responses are deliberately corrupted, so the expected score loss is proportional to $r$. Notably, HumanEval+ is almost immune to poisoning (DC$_\text{HE+} \leq 0.012$), likely because test cases catch the corrupted solutions during evaluation, and the teacher's code generation is robust enough that even ``deliberately wrong'' code often still passes.

\paragraph{Token limit $L{=}1024$ is cheapest.} Token truncation at $L{=}1024$ (A10) has the lowest DC of any defense ($\overline{\text{DC}} = 0.014$), with zero cost on HumanEval+ and MT-Bench and only a 4\% reduction on MATH-500. However, its DE is correspondingly negligible ($\overline{\text{DE}} = 0.985$), illustrating the fundamental tension: low-cost defenses tend to be low-effectiveness.

\begin{figure}[t]
\centering
\begin{tikzpicture}
\begin{axis}[
  width=\columnwidth, height=6cm,
  xlabel={$\overline{\text{DC}}$ (user cost $\rightarrow$)},
  ylabel={$\overline{\text{DE}}$ (distillation quality $\rightarrow$)},
  xmin=-0.02, xmax=0.36,
  ymin=0.75, ymax=1.06,
  grid=major, grid style={gray!20},
  every axis plot/.append style={only marks, mark size=4pt},
  legend style={at={(0.02,0.02)}, anchor=south west, font=\small},
  clip=false,
]
\draw[gray!40, dashed, thick] (axis cs:0.05,0.75) -- (axis cs:0.05,0.85) -- (axis cs:-0.02,0.85);
\node[gray!60, font=\scriptsize, anchor=north west] at (axis cs:-0.01,0.84) {ideal};

\addplot[color=pertcolor, mark=*] coordinates {(0.030,0.990) (0.034,0.978) (0.070,1.012)};
\addplot[color=poiscolor, mark=square*] coordinates {(0.017,0.979) (0.052,0.962) (0.098,0.980)};
\addplot[color=thrtcolor, mark=triangle*] coordinates {(0.311,0.811) (0.046,0.986) (0.014,0.985)};

\legend{Perturbation, Poisoning, Throttling}

\node[font=\tiny, anchor=south west] at (axis cs:0.032,0.992) {A02};
\node[font=\tiny, anchor=south east] at (axis cs:0.030,0.976) {A03};
\node[font=\tiny, anchor=south west] at (axis cs:0.072,1.014) {A04};
\node[font=\tiny, anchor=south east] at (axis cs:0.015,0.977) {A05};
\node[font=\tiny, anchor=north west] at (axis cs:0.054,0.960) {A06};
\node[font=\tiny, anchor=south west] at (axis cs:0.100,0.982) {A07};
\node[font=\tiny, anchor=west] at (axis cs:0.320,0.811) {A08};
\node[font=\tiny, anchor=south west] at (axis cs:0.048,0.988) {A09};
\node[font=\tiny, anchor=south east] at (axis cs:0.012,0.983) {A10};
\end{axis}
\end{tikzpicture}
\caption{DE--DC trade-off for all nine defenses. The ideal defense (low DE, low DC) would occupy the bottom-left corner. Instead, all defenses either cluster near the top (high DE $\approx$ no protection) with low DC, or achieve lower DE only at substantial DC cost (A08). No defense reaches the ideal region.}
\label{fig:de_dc}
\end{figure}
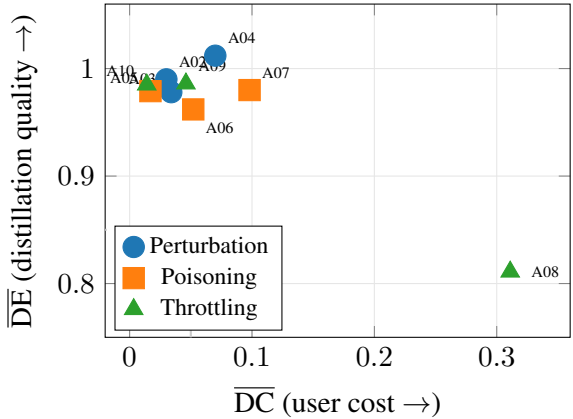

\paragraph{No defense achieves low DE and low DC simultaneously.} This is the key finding. Figure~\ref{fig:de_dc} plots $\overline{\text{DE}}$ against $\overline{\text{DC}}$ for all nine defenses. The ideal defense would occupy the bottom-left corner (low DE, low DC), but all defenses cluster along an unfavorable frontier: defenses with low $\overline{\text{DE}}$ (stronger protection) invariably have high $\overline{\text{DC}}$ (greater user cost), while low-cost defenses provide little protection. The only defense with $\overline{\text{DE}} < 0.95$ is CoT removal, which pays a DC of 0.311. This reinforces our central finding that output-level defenses face a fundamental trade-off between protection and utility.

\section{Discussion}
\label{sec:discussion}

\subsection{Why Output-Level Defenses Fail}

Our results point to a fundamental tension in output-level defense design. The defender faces a dual-use dilemma: any API output that is useful to legitimate users is also useful for distillation. This is because knowledge distillation extracts the \emph{functional mapping} from prompts to responses---the same mapping that provides value to users.

This tension manifests differently for each defense category:
\begin{itemize}[leftmargin=*]
  \item \textbf{Perturbation} preserves both user value and distillation value---it cannot degrade one without degrading the other.
  \item \textbf{Poisoning} degrades both, but distillation quality degrades more slowly than user experience because task-specific capabilities are robust to moderate noise.
  \item \textbf{Throttling} can selectively degrade distillation value for reasoning tasks (by removing CoT traces), but at the cost of also degrading user value for the same tasks.
\end{itemize}

The only escape from this tension appears to be defenses that operate outside the output stream: watermarking modifies the \emph{distribution} of outputs rather than individual responses, query detection operates on \emph{inputs}, and model-level defenses modify the model itself. These approaches can potentially provide protection without proportionally degrading user value.

\subsection{Implications for Defense-Aware Attackers}

Our evaluation considers only the naive attacker. Stronger attackers could potentially overcome even the effective defenses we identified:
\begin{itemize}[leftmargin=*]
  \item \textbf{Against poisoning}: A filtering attacker could verify answers using external solvers or test cases, discarding poisoned responses.
  \item \textbf{Against throttling}: A student-CoT attacker could train the student to generate its own chain-of-thought reasoning (using the correct final answers as supervision), partially recovering the lost reasoning capability.
  \item \textbf{Defense identification}: An attacker could probe the API to identify which defense is deployed (e.g., by comparing responses to repeated queries for perturbation detection, or checking for reasoning trace presence for throttling detection), then adapt their strategy accordingly.
\end{itemize}

These adaptive attack strategies suggest that our DE measurements are upper bounds on the protection that output-level defenses can provide. Evaluating defense-aware attackers is a key direction for future work.

\subsection{Ethical Considerations}

Our work systematically evaluates the effectiveness of defenses against knowledge distillation, which inherently involves understanding attack strategies. We believe this research is beneficial because:
\begin{enumerate}[leftmargin=*]
  \item \textbf{Informed defense design}: By demonstrating that current defenses fail, we help providers avoid deploying ineffective measures and redirect effort toward more promising approaches.
  \item \textbf{No novel attack capability}: The naive attacker we evaluate is trivially simple and well-known in the community. Our contribution is on the \emph{defense evaluation} side, not the attack side.
  \item \textbf{Responsible framing}: We present results as a call for better defenses, not as a guide for conducting attacks.
\end{enumerate}

We acknowledge that a complete evaluation of defense-aware attacks (planned for future work) requires more careful ethical consideration, as it involves developing and publishing novel attack strategies.

\subsection{Limitations}
\label{sec:limitations}

Our evaluation has several limitations that qualify the generalizability of our findings:

\begin{itemize}[leftmargin=*]
  \item \textbf{Single teacher--student pair.} We evaluate only one distillation setting: Qwen3-14B $\to$ Qwen2.5-7B-Instruct, both from the Qwen family. Same-family distillation may be easier than cross-family settings (e.g., GPT-4 $\to$ LLaMA), so our DE results may represent a best-case scenario for the attacker. Defense effectiveness could be higher in cross-family settings where the student's architecture is less aligned with the teacher's output distribution.
  \item \textbf{Naive attacker only.} We evaluate only the simplest attacker strategy: querying once per prompt with no filtering or post-processing. Stronger attackers---consensus voting, answer verification, or defense-aware adaptation---would likely further reduce defense effectiveness, making our DE results an upper bound on defense protection.
  \item \textbf{Fixed training data scale.} All experiments use 10{,}000 training samples. Defense effectiveness may vary with dataset size: defenses like poisoning might be more or less effective at different scales.
  \item \textbf{Judge model bias.} We use Qwen2.5-7B-Instruct as the MT-Bench judge, which is the same model as the student base. This introduces a potential self-preference bias: the judge may systematically favor responses from models in the same family. While this bias is consistent across all experiments (and thus does not affect relative comparisons), it may inflate absolute MT-Bench scores.
\end{itemize}

\section{Related Work}
\label{sec:related}

\paragraph{Model stealing and extraction.}
The threat of model extraction via prediction APIs was first formalized by \citet{tramer2016stealing}, who demonstrated that ML models (logistic regression, decision trees, neural networks) could be functionally replicated through targeted queries to prediction APIs. \citet{krishna2020thieves} extended this line of work to NLP, demonstrating that BERT-based models fine-tuned for classification and question answering could be extracted through API querying, with the extracted models achieving near-parity performance. More recently, \citet{carlini2024stealing} demonstrated that parts of production language models (specifically, projection layers) could be extracted through carefully crafted API queries, blurring the line between functional replication and architectural extraction.

Our work differs from these in focusing specifically on \emph{knowledge distillation} as the extraction mechanism---training a student model on teacher outputs---rather than attempting to recover model weights or architecture. This distillation threat is arguably more practical for LLMs because it requires no knowledge of the target model's architecture and can be performed with standard fine-tuning tools.

\paragraph{LLM distillation in practice.}
The practical threat of API distillation has been demonstrated by numerous projects. Stanford Alpaca \citep{taori2023alpaca} fine-tuned LLaMA-7B \citep{touvron2023llama} on 52K instruction-output pairs generated by GPT-3.5, producing a model that exhibited instruction-following capabilities comparable to the teacher. Vicuna \citep{chiang2023vicuna} took a similar approach using 70K conversations from ShareGPT, achieving what the authors described as ``90\% of ChatGPT quality.'' WizardLM \citep{xu2023wizardlm} introduced instruction evolution to increase training data complexity, while Orca \citep{mukherjee2023orca} leveraged chain-of-thought explanation traces from GPT-4 to teach reasoning processes rather than just answers. The phi series \citep{gunasekar2023textbooks} demonstrated that carefully curated synthetic training data could produce surprisingly capable small models. These works collectively demonstrate the severity and accessibility of the distillation threat but do not evaluate or propose defensive countermeasures.

\paragraph{Knowledge distillation methods.}
Classical knowledge distillation \citep{hinton2015distilling} transfers knowledge from a teacher to a student through soft probability distributions (logits). Recent work has adapted this paradigm to LLMs, though API-based distillation is limited to output-level supervision (no logit access). MiniLLM \citep{gu2024minillm} uses reverse KL divergence for more effective sequence-level distillation. Distilling Step-by-Step \citep{hsieh2023distilling} demonstrates that chain-of-thought rationales provide a powerful additional supervisory signal, enabling smaller students to outperform larger models trained without rationales. The instruction-tuning paradigm \citep{peng2023instruction} can itself be viewed as a form of distillation from stronger to weaker models. Our work evaluates defenses from the \emph{provider's perspective}---how to prevent this knowledge transfer---rather than optimizing the distillation process itself.

\paragraph{Defenses against model extraction.}
Watermarking \citep{kirchenbauer2023watermark,aaronson2023watermarking} represents a detection-based defense that embeds statistical patterns in model outputs, enabling post-hoc identification of distilled models. Unlike our output-level defenses, watermarking aims to \emph{detect} rather than \emph{prevent} distillation. Differential privacy \citep{dwork2006calibrating} provides formal guarantees on information leakage but requires adding noise proportional to the sensitivity of the output, which is impractically large for LLM text generation. Our output-level defenses (perturbation, poisoning, throttling) represent a more practical but less principled approach. To our knowledge, no prior work has systematically evaluated and compared these output-level defense strategies in a controlled experimental setting with modern LLMs, which is the primary contribution of our work.

\section{Conclusion}
\label{sec:conclusion}

We presented DistillGuard, a framework for evaluating output-level defenses against LLM knowledge distillation. Our evaluation of nine defense configurations across three categories reveals a sobering finding: most output-level defenses are ineffective against even naive distillation attacks. Paraphrasing-based perturbation provides no meaningful protection, data poisoning primarily degrades conversational quality without protecting core capabilities, and only chain-of-thought removal substantially impairs reasoning---while leaving other capabilities intact.

Our distillation cost (DC) analysis further reveals that the only effective defense---CoT removal---imposes a severe cost on legitimate users ($\overline{\text{DC}} = 0.311$), with the teacher's own math accuracy dropping from 78.4\% to 12.6\%. No defense achieves simultaneously low DE and low DC, confirming a fundamental trade-off in output-level defense design.

These results suggest that the current landscape of output-level defenses is inadequate for protecting proprietary LLM capabilities. The perturbation limitation we identify---that semantic-preserving transformations empirically preserve distillation value---points to a key constraint of this defense paradigm. Future work should explore: (1)~defense-aware adaptive attacks that may further erode the limited protection these defenses provide; (2)~defense combinations that may achieve broader coverage across capability dimensions; and (3)~fundamentally different defense paradigms such as watermarking and fingerprinting that do not rely on degrading output quality. Our framework's pipeline can also be extended to iterative collection-training loops for evaluating active learning attacks and to cross-family distillation settings for broader generalizability.


\appendix

\section{Defense Implementation Details}
\label{app:defense}

\paragraph{Paraphrasing perturbation.}
The paraphraser receives the teacher's response along with a system prompt parameterized by $\alpha$. At $\alpha = 0.3$, the prompt instructs: ``Lightly rephrase the following response, preserving all technical content, structure, and key phrases. Only change minor wording.'' At $\alpha = 0.7$: ``Substantially rephrase the response. Change sentence structure, reorganize content, and use different vocabulary while preserving correctness.'' At $\alpha = 1.0$: ``Completely rewrite the response in your own words. You may reorganize, restructure, and rephrase freely as long as the answer remains correct.'' The paraphraser operates with a maximum context length of 16{,}384 tokens and temperature 0 for deterministic output.

\paragraph{Response corruption.}
For each prompt selected for poisoning (with probability $r$), the teacher model is re-queried with an adversarial system prompt: ``Solve the following problem, but deliberately arrive at an incorrect answer. Make your solution look plausible and well-reasoned, but ensure the final answer is wrong.'' This produces responses that are stylistically consistent with genuine ones but contain systematic errors. The poisoned and clean responses are shuffled randomly.

\paragraph{CoT removal.}
We implement CoT removal by extracting only the final answer from the teacher's response. For math problems, we search for answer markers (e.g., ``the answer is'', ``$=$'', boxed expressions) and extract the final numerical or algebraic answer. For code, we extract only the function body. For open-ended responses, we take only the concluding paragraph.

\paragraph{Token truncation.}
Truncation is applied at the token level using the student model's tokenizer. The teacher's response is tokenized, truncated to $L$ tokens, and detokenized. We do not attempt to truncate at sentence boundaries, as this would conflate truncation with content selection.

\section{Baseline Model Quality}
\label{app:baseline}

For reference, the base Qwen2.5-7B-Instruct model (before any distillation) achieves MATH-500: 68.4\%, HumanEval+: 94.5\%, MT-Bench: 8.04 on our evaluation setup. The distilled baseline (A01: 67.8\%, 91.5\%, 8.20) is comparable, with a slight improvement on MT-Bench and slight declines on MATH and HumanEval+. This indicates that distillation from Qwen3-14B provides marginal benefit over the already-capable base model, consistent with the same-family setting where teacher and student share pretraining.

The CoT-removal student (A08: 31.4\%) performs substantially \emph{worse} than the undistilled base model (68.4\%) on MATH-500. This confirms that training on answer-only data is actively harmful for reasoning---the student not only fails to gain capability from distillation but loses capability it originally had.

\section{Reproducibility}
\label{app:repro}

All experiments use a single NVIDIA GPU with deterministic operations enabled (seed 42). The full pipeline---teacher generation, defense application, student training, and evaluation across all three benchmarks---takes approximately 15 hours per experiment. Total compute for the ten experiments in this paper is approximately 150 GPU-hours. Code, configuration files, and evaluation scripts are organized in a reproducible pipeline and will be released upon publication.

\end{document}